\documentclass[]{aa}
\usepackage{graphicx}
\begin{document}

\title{Theoretical isochrones for the $\Delta a$ photometric system}
\author{A.~Claret\inst{1}, E.~Paunzen\inst{2,3}, H.M.~Maitzen\inst{2}}

\mail{claret@iaa.es}

\institute{Instituto de Astrof\'\i sica de Andaluc\'\i a, CSIC, Apartado 3004, 
                   E-18080 Granada, Spain
\and       Institut f\"ur Astronomie der Universit\"at Wien,
           T\"urkenschanzstr. 17, A-1180 Wien, Austria
\and       Zentraler Informatikdienst der Universit\"at Wien,
           Universit\"atsstr. 7, A-1010 Wien, Austria}

\date{Received 2003; accepted 2003}
\titlerunning{Theoretical isochrones for $\Delta a$}{}
\abstract{We have calculated theoretical isochrones for the
photometric $\Delta a$ system to derive astrophysical parameters such
as the age, reddening and distance modulus for open clusters. The
$\Delta a$ system samples the flux depression at 520\,nm which is
highly efficient to detect chemically peculiar (CP) objects of the
upper main sequence. The evolutionary status of CP stars is still a
matter of debate and very important to test, for example, the dynamo
and diffusion theories. In fact, the dynamo or fossil origin of
the magnetic fields present in this kind of stars it still not clear.
Using the stellar evolutionary models by Claret (1995), a grid of
isochrones with different initial chemical compositions for the
$\Delta a$ system was generated.  The published data of 23 open
clusters were used to fit these isochrones with astrophysical
parameters (age, reddening and distance modulus) from the
literature. As an additional test, isochrones with the same
parameters for Johnson $UBV$ data of these open clusters were 
also considered. The fits show a good agreement between the
observations and the theoretical grid. We find that the accuracy of
fitting isochrones to $\Delta a$ data without the knowledge of the
cluster parameters is between 5 and 15\,\%.
}
\maketitle

\section{Introduction}

The chemically peculiar (CP) stars of the upper main sequence have
been targets for astrophysical studies since their discovery by
the American astronomer Antonia Maury (1897).  Most of this interval
was devoted to the detection of peculiar features in their spectra and
photometric behaviour.  The main characteristics of the classical CP
stars are: peculiar and often variable line strengths, quadrature of
line variability with radial velocity changes, photometric variability
with the same periodicity and coincidence of extrema. Overabundances
of several orders of magnitude compared to the Sun were derived for
heavy elements such as Silicon, Chromium, Strontium and Europium.

The discovery by Babcock (1947) of a global dipolar magnetic
field in the star 78 Virginis was followed by a catalog of similar
stars (Babcock 1958) in which also the variability of the field
strength in many CP stars, including even a reversal of magnetic
polarity was discovered leading the Oblique Rotator concept of slowly
rotating stars with non-coincidence of the magnetic and rotational
axes (Babcock 1949).  Due to the chemical abundance concentrations at
the magnetic poles spectral and the related photometric
variabilities are also easily understood, as well as radial velocity
variations of appearing and receding patches on the stellar surface.

The prerequisite for investigating larger samples of CP stars
(including the generally fainter open cluster members) is unambiguous
detection. Looking into catalogues of CP stars, especially of the
magnetic ones, it immediately becomes obvious that there are many
discrepancies at classification dispersions.  Even the short list of
peculiar stars identified by Maury (1897) contains one object which is
classified as erroneous in the catalogue of Renson (1991).

The reasons for discrepant peculiarity assessments are to be found in
the differences of personal pattern recognition among different
classifiers, differences of (mainly photographic) observing material
(density of spectrograms, widening of spectra, dispersion, focussing),
seeing conditions (for objective prism spectra), and intrinsic
variability of peculiar spectral features (e.g. silicon lines).

Photometry has shown a way out of this dilemma, especially through the
discovery of characteristic broad band absorption features; the
most suitable being located around 520\,nm.  Two decades ago, Maitzen
(1976) introduced a three filter photometric system which samples the
depth of this flux depression by comparing the flux at the center
(522\,nm $g_{\rm 2}$), with the adjacent regions (500\,nm, $g_{\rm 1}$
and 550\,nm, $y$) using a band-width of 13\,nm.

In this paper we have used the published CCD photometry of 23 open
clusters (Bayer et al. 2000, Maitzen et al. 2001, Paunzen \& Maitzen
2001, 2002 and Paunzen et al. 2002, 2003) together with theoretical
isochrones for the $\Delta a$ system to test the capability of this
system to derive the reddening, age and distance for these open
clusters. The isochrones are based on the evolutionary stellar models
by Claret (1995). We find a very good agreement of the fitted
parameters for the $\Delta a$ system compared to those from the
literature for e.g. the Johnson $UBV$ system.

\begin{figure}
\begin{center}
\includegraphics[width=80mm]{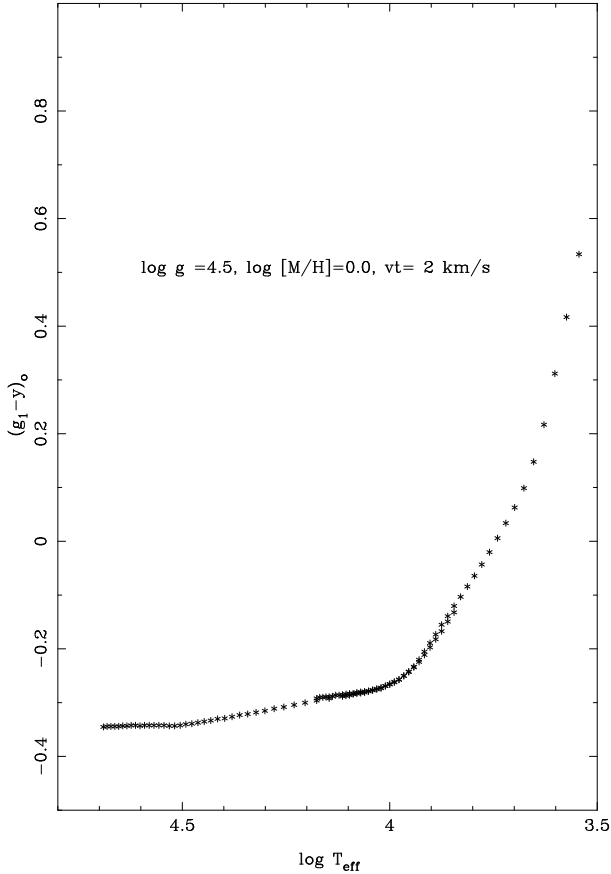}
\caption{Theoretical relationship between ($g_1-y$)$_{0}$ and the effective temperature for
models with solar abundance, vt\,=\,2\,kms$^{-1}$, {\it l}/H$_p$\,=\,1.25 and log\,$g$\,=\,4.5
dex. For models with T$_{eff} \le$ 8500 K, an alternative theory of convection was adopted (see 
text). }
\end{center}
\end{figure}

\begin{figure}
\begin{center}
\includegraphics[width=80mm]{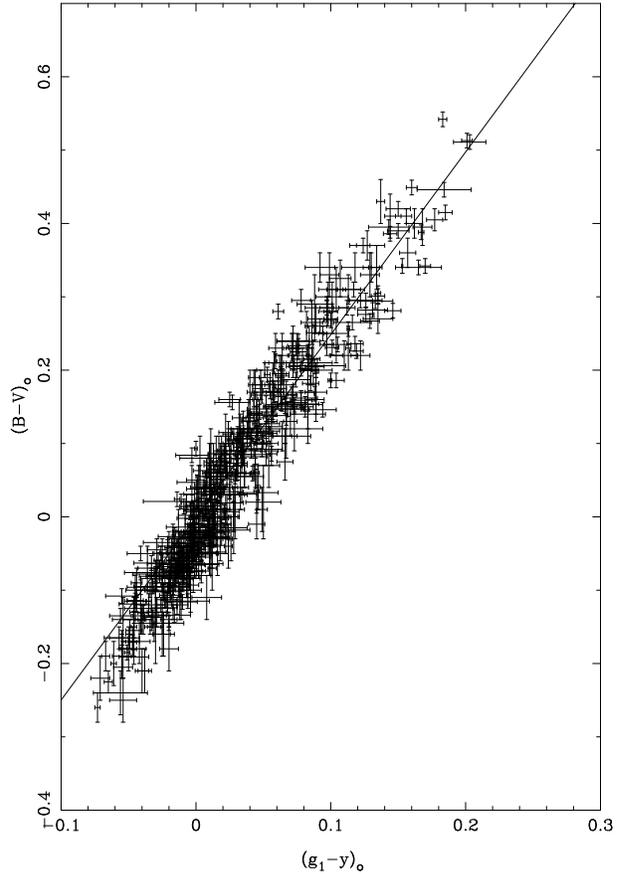}
\caption{The empirical ($g_1-y$)$_o$ and ($B-V$)$_o$ data and the respective 
theoretical predictions (continuous line). The theoretical value was obtained 
by averaging the whole grid of stellar atmosphere models. }
\end{center}
\end{figure}

\begin{figure}
\begin{center}
\includegraphics[width=80mm]{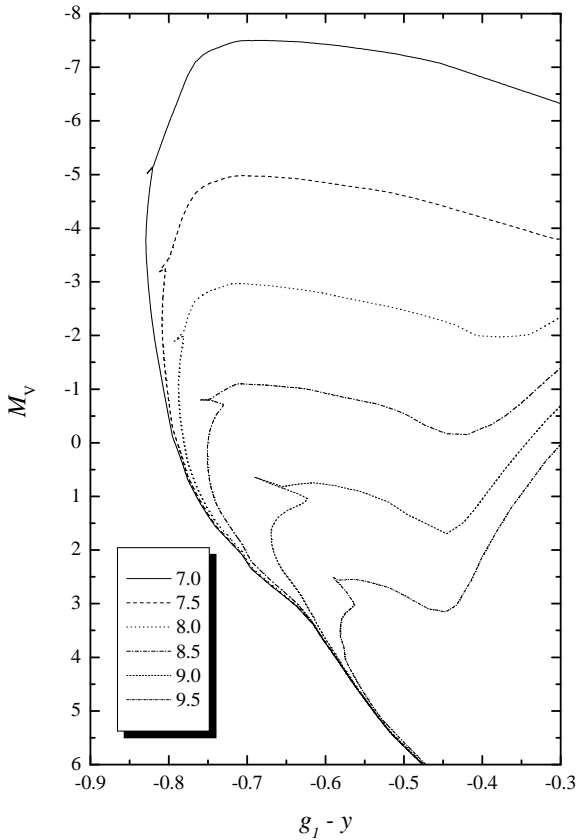}
\caption{Theoretical isochrones for different ages and solar abundance for the 
evolutionary models described in Sect. 3}
\end{center}
\end{figure}

\begin{table*}[t]
\begin{center}
\caption{23 published open clusters with $\Delta a$ CCD photometry
taken from Bayer et al. (2000), 
Maitzen et al. (2001), Paunzen \& Maitzen (2001, 2002) and Paunzen et al. (2002, 2003). 
The
values for the reddening, age and distance modulus are from the  same sources. 
The metallicity for all clusters is (within the errors) solar according to
WEBDA and Lyng\aa\, (1987).}
\label{ucl}
\begin{tabular}{ccclc|ccclc}
\hline
\hline
Cluster & log\,$t$ & $m_V - M_V$ & \multicolumn{1}{c}{E($B-V$)} & $N_{Stars}$ &
Cluster & log\,$t$ & $m_V - M_V$ & \multicolumn{1}{c}{E($B-V$)} & $N_{Stars}$ \\
\hline
Collinder 272 & 7.20 & 13.06 & 0.47 & 111 & NGC 6134 & 8.97 & 11.07 & 0.40 & 102 \\
Lyng{\aa}\, 14 & 6.71 & 14.29 & 1.48 (var) & 53 & NGC 6192 & 7.95 & 13.00 & 0.68 & 98 \\
Melotte 105 & 8.30 & 12.70 & 0.36 & 114 & NGC 6204 & 7.60 & 11.50 & 0.43 & 319 \\
NGC 2099 & 8.50 & 11.67 & 0.30 & 41 & NGC 6208 & 9.00 & 10.54 & 0.18 & 41 \\
NGC 2169 & 7.70 & 10.57 & 0.12 & 13 & NGC 6250 & 6.50 & 11.18 & 0.38 & 48 \\
NGC 2439 & 7.20 & 14.23 & 0.41 & 113 & NGC 6396 & 7.20 & 13.34 & 0.96 (var) & 105 \\
NGC 2489 & 8.45 & 12.00 & 0.40 & 59 & NGC 6451 & 8.13 & 13.74 & 0.67 & 146 \\
NGC 2567 & 8.43 & 11.40 & 0.13 & 50 & NGC 6611 & 6.90 & 13.72 & 0.86 (var)&  79\\
NGC 2658 & 9.10 & 11.67 & 0.04 & 55 & NGC 6705 & 8.30 & 12.73 & 0.43 & 312 \\
NGC 3114 & 8.10 & 10.02 & 0.07 & 271 & NGC 6756 & 8.10 & 14.67 & 0.80 & 65 \\
NGC 3960 & 8.80 & 12.74 & 0.30 & 93 & Pismis 20 & 6.60 & 15.30 & 1.24 (var)& 238 \\
NGC 5281 & 7.10 & 11.35 & 0.26 & 30 \\
\hline
\end{tabular}
\end{center}
\end{table*}

\begin{figure*}
\begin{center}
\includegraphics[width=115mm,angle=270]{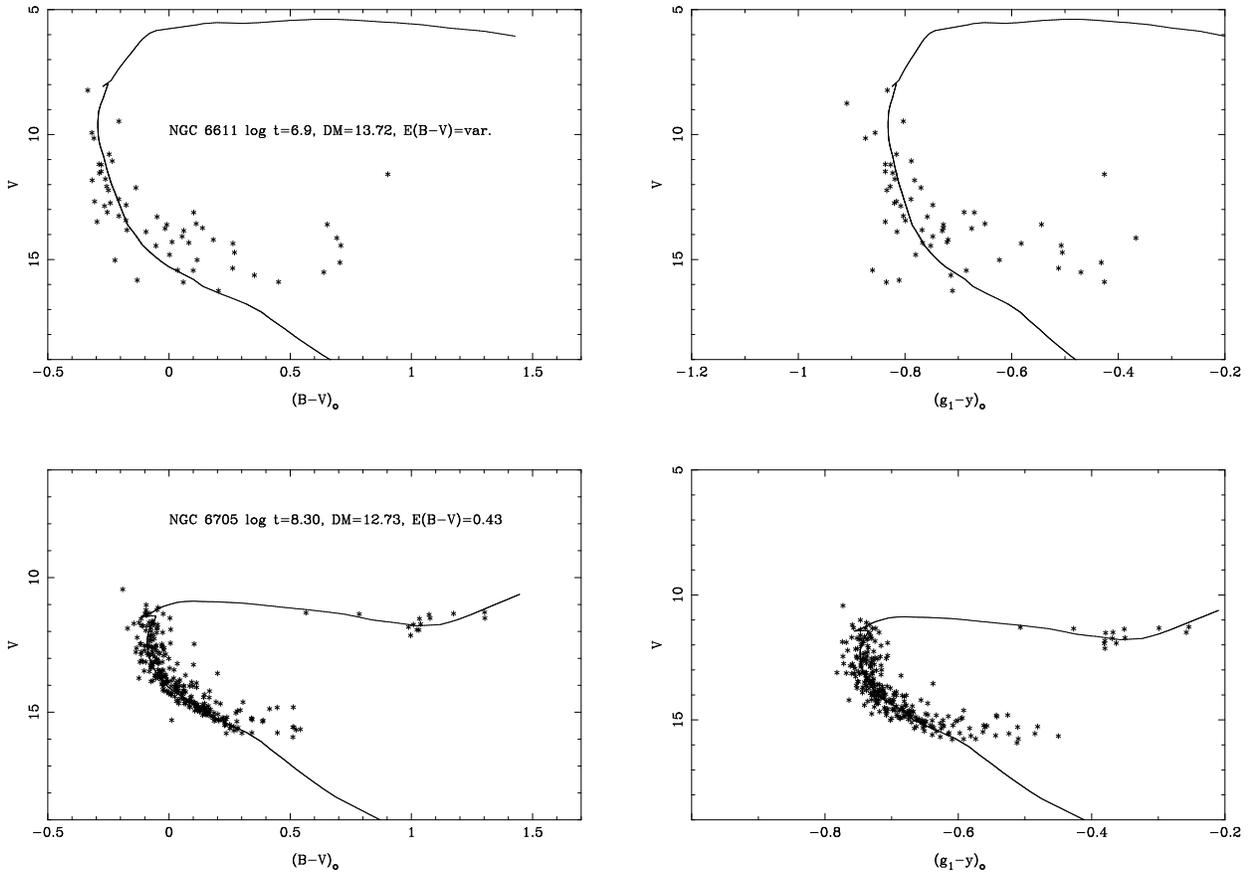}
\caption[]{Isochrones for NGC 6611 and NGC 6705 with the parameters listed
in Table 1.}
\end{center}
\end{figure*}

\section{Motivation and aim}

The $\Delta a$ system is based on the three narrow band filters $g_1, g_2$ and $y$. 
The respective index measuring the 520\,nm depression was introduced as:
$$ a = g_{\rm 2} - (g_{\rm 1} + y)/2 $$
Since this quantity is slightly dependent on temperature (increasing
towards lower temperatures), the intrinsic peculiarity index
had to be defined as
$$ \Delta a = a - a_{\rm 0}[(b - y); (B - V); (g_{\rm 1} - y)] $$
i.e. as the difference between the individual $a$-value and the
$a$-value of non-peculiar stars of the same colour (the locus of the
$a_{\rm 0}$-values has been called the normality line). It
has been shown that virtually all peculiar stars with magnetic fields
(CP2 stars) have positive $\Delta a$-values up to 0.075\,mag whereas
Be/Ae and $\lambda$ Bootis stars exhibit significant negative
ones. Extreme cases of the CP1 and CP3 group may exhibit marginally
peculiar positive $\Delta a$ values (Vogt et al. 1998).

The index $(g_{\rm 1} - y)$ shows an excellent correlation with
$(B-V)$ as well as $(b-y)$ and can be used as an indicator for the
effective temperature.  The main result of detecting peculiar stars is
supplemented by a very accurate color-magnitude diagram ($y$ or $V$
versus $(g_1-y)$) which can be used to determine the reddening,
distance and age of an open cluster.

The evolutionary status of the CP stars has been especially
controversial. Oetken (1984) concluded that the CP2 (magnetic CP
stars) phenomenon appears at the late stages of the main sequence
evolution. Later, Hubrig et al. (2000) found that the distribution of
CP2 stars of masses below 3\,$M_{\odot}$ in the
Hertzsprung-Russell-diagram differs from that of the presumably
normal-type stars in the same temperature range at a high level of
significance. Magnetic stars are concentrated toward the center of the
main sequence band. In particular, Hubrig et al. (2000) found
that measurable magnetic fields appear only in stars which have
already completed at least approximately 30\% of their main sequence
life-time. No clear picture emerges as to the possible evolution of
the magnetic field across the main sequence. Weak hints of some
(loose) relations between magnetic field strength and other stellar
parameters are found: stars with shorter rotational periods tend to
have stronger fields, as do higher temperature and higher mass
stars. No correlation between the rotation period and the fraction of
the main sequence life time completed was observed, indicating that
the slow rotation in these stars must already have been achieved
before they became observably magnetic. A marginal trend of the
magnetic flux to be lower in more slowly rotating stars may possibly
be seen as suggesting a dynamo origin for the magnetic field.

The results of the Hipparcos mission on the other hand do not support
the mentioned above findings. G\'omez et al. (1998) presented the
Hertzsprung-Russell-diagram of about 1000 CP stars in the solar
neighbourhood using astrometric data from the Hipparcos
satellite as well as photometric and radial velocity data. Most CP
stars lie on the main sequence occupying the whole width of it (about
2 mag), just like normal stars in the same range of spectral
types. P{\"o}hnl et al. (2003) present further evidence that the CP2
stars already occur at very early stages of stellar evolution,
significantly before they reach 30\,\% of their life-time on the main
sequence. P{\"o}hnl et al. (2003) investigated four young open
clusters with known ages and accurate distances (error\,$<$\,10\%),
including CP2 members using the measurements and calibrations of the
Geneva 7-color photometric systems to derive effective temperatures
and luminosities. Taking into account the overall metallicity of the
individual clusters, isochrones and evolutionary tracks were used to
estimate ages and masses for the individual objects. The derived ages
(between 10 and 140\,Myr) were well in line with those of the
corresponding clusters.

The finding of CP stars in open clusters is essential to put further
constraints on models dealing with dynamo theories, angular momentum
loss during the pre- as well as main sequence and evolutionary
calculations for such objects.

The age, reddening and distance of open clusters is often
controversial (see Paunzen \& Maitzen 2002 for the case of NGC 6451)
or not even defined at all. We have therefore calculated isochrones
for the $\Delta a$ system to perform such an analysis.

\section{Models, data selection and isochrone fitting} \label{models}

The stellar models used to calculate the isochrones have been
described in more detail by Claret (1995). We will give here a short
overview of the input physics. The chemical composition is (X, Z) =
(0.70, 0.02) though different combinations of X and Z can also be
used following the metallicity indicator of each cluster.  The
equation of state takes into account partial ionization through Saha's
approximation, the pressure of gas and radiation as well as the
equations for degenerate electrons. For less massive stars, a special
treatment of the equation of state was adopted through the CEFF
package (D{\"a}ppen 1994\,$-$\,2000).  Radiative opacities were
computed with the OPAL code.  The mixing length theory was used to
describe convection and moderate overshooting with $\alpha_{ov}$
 = 0.2 was considered for convective cores. The models take into
account mass loss during the main sequence as well as during the
red giant phase.

The filter transmission functions of $g_1$ and $y$ are the same as
those used by Kupka et al. (2003). We have performed synthetic colour
calculations, using the properties of the mentioned filters, in order
to establish the connection between observed and theoretical
quantities. A similar procedure as described by Castelli (1999) and
Kupka et al.  (2003) was adopted. The original ATLAS9 models (see
Kurucz 1993), with a microturbulence velocity of 2 kms$^{-1}$, $-$5.0
$\le$ log [M/H] $\le$ +1 and mixing-length parameter $\alpha$ = 1.25
were the basic tools to derive the synthetic colours. For T$_
{eff} \le$ 8500\,K, we adopted the calculations by Kupka et al. (2003)
due to problems detected in the convection model, precisely in the
specific intensities (e.g. Claret et al. 1995). These models were
computed adopting the Canuto \& Mazzitelli (1991) prescription. The
implementation of this alternative theory of transport of energy by
convection does not significantly affect the hotter models, as
expected. In fact, the differences in the respective colour indices
are of the order of 1 to 3 mmag, as described and calculated by Kupka
et al. (2003).  As an example of synthetic colour calculation Fig. 1
shows the dependence of ($g_1-y$)$_o$ with the effective temperature
for models with log\,$g$\,=\,4.5 dex.  Similar results are obtained
for different values of log\,$g$, microturbulent velocities and/or
metallicity.  Figure 2 shows a comparison between empirical data and
theoretical colour indices. The continuous line indicates the average
of all theoretical models (in log\,$g$).  The zero-point was corrected
by adding 0.31 to the theoretical ($g_1-y$) values.  The final
isochrones, considering the adequate values of log\,$g$ for each class
of luminosity, are shown in Fig. 3 for different ages from
log\,$t$\,=\,7.0 to 9.5, respectively.

Kupka et al. (2003) have shown that metallicities different from solar
values shift the normality line by about 3 to 6 mmag for [Z] =
$\pm$0.5\,dex. Such an effect is, in general, a factor of two smaller
than the intrinsic measurement errors and not
detectable. Nevertheless, we have searched through the open cluster 
database WEBDA 
(http://obswww.unige.ch/webda/) and Lyng\aa\, (1987) for available
metallicities of the open clusters listed in Table\,\ref{ucl}. For none
of the investigated clusters, a value different than solar was
found within the given error. We therefore expect that all programme
clusters have solar metallicity.

We have taken the published $\Delta a$ CCD photometry together with
the Johnson $UBV$ ones of 23 open clusters (Bayer et al. 2000, Maitzen
et al. 2001, Paunzen \& Maitzen 2001, 2002 and Paunzen et al. 2002,
2003) to test these isochrones. These open clusters have widely
different ages and reddening which makes them excellent test
cases (Table 1).

The isochrone fitting was performed in two steps. First, we 
dereddened the ($g_1-y$) values according $E(g_1-y) = 0.4\cdot
E(B-V)$ (Maitzen 1993) and calculated an individual isochrone
according to the ages (taken from the literature) listed in Table
1. The data were then plotted with the appropriate distance
modulus. In order to test the parameters from the literature, the same
procedure was performed for Johnson $UBV$ colors. Figure 4 shows the
examples of NGC 6611 and NGC 6705 for both photometric systems,
respectively. The isochrones fit, in general, the data very well.

As second step, the $\Delta a$ data were fitted to the isochrones
without an a-priori knowledge of the reddening, age and distance
modulus. Here we face the same problems and error sources as for the
classical photometric systems since no colour-colour diagram is
available. Nevertheless, we were able to reproduce the parameters from
the literature with an accuracy between 5 and 15\,\% depending on the
age, available giant members and the presence of differential
reddening.

\section{Conclusions}

We have investigated the capability of theoretical isochrones for the
photometric $\Delta a$ system to derive astrophysical parameters such
as the age, reddening and distance modulus for open clusters. The
$\Delta a$ system is highly efficient to detect chemically peculiar
objects of the upper main sequence by sampling the flux depression at
520\,nm.

Using the stellar evolutionary models by Claret (1995) and the well
established filter transmission functions, a grid of isochrones with
solar abundances was calculated.

As a test, the published data of 23 open clusters were used to fit
these isochrones with parameters from the literature. Furthermore, the
appropriate isochrones with the same parameters for the Johnson $UBV$
photometric system were considered. The fits show an excellent
agreement between the observations and the theoretical grid.

In addition, we have fitted the observational data without the
knowledge of the age, reddening and distance modulus yielding an
accuracy of 5 to 15\% depending on the well known error sources of
such a method.

\begin{acknowledgements}
EP acknowledges partial support by the Fonds zur F\"orderung der
wissenschaftlichen Forschung, project P14984. The Spanish DGYCIT
(PB98-0499) is gratefully acknowledged for its support during the
development of this work. We are grateful to F. Leone and
B. Willems, who have helped
to improve the final version of the paper.  Use was made of the SIMBAD
database, operated at CDS, Strasbourg, France and the WEBDA database,
operated at the Institute of Astronomy of the University of Lausanne.
\end{acknowledgements}

\end{document}